\documentclass[aps,superscriptaddress,preprint,nofootinbib,floatfix]{revtex4}
\usepackage{bm}
\usepackage{dcolumn}
\usepackage{epsfig}
\usepackage{pstricks}
\def \ugev {\mathrm{GeV}}

\def \eq#1{Eq.~(\ref{#1})}

\def \fig#1{Fig.~\ref{#1}}
\def \figs#1#2{Figs.~\ref{#1}--\ref{#2}}
\def \rf{Ref.~\cite}

\begin{document}
%
%
%
\title{Charged and neutral pion production in the S-matrix approach}
\author{V.~Malafaia}
\affiliation{Centro de  F\'isica Te\'orica de Part\'iculas,
Instituto Superior T\'ecnico}
%
\author{M.T.~Pe\~na}
\affiliation{Centro de  F\'isica Te\'orica de Part\'iculas,
Instituto Superior T\'ecnico} \affiliation{Department of Physics,
Instituto Superior T\'ecnico, Av. Rovisco Pais, 1049-001 Lisbon,
Portugal }
%
\author{Ch.~Elster}
\affiliation{Institute of Nuclear and Particle Physics, and
Department of Physics and Astronomy, Ohio University, Athens, OH
45701}
%
\author{J.~Adam, Jr.}
\affiliation{Nuclear Physics Institute, \v{R}e\v{z} near Prague,
    CZ-25068, Czech Republic}
\date{\today}
%
%
\begin{abstract}
The S-matrix approach is used to calculate charged as well as
neutral pion production reactions from $NN$ scattering, with the
same set of underlying processes and interactions.
The chiral perturbation theory ($\chi$PT) $\pi$N scattering
amplitude is used. For the nucleon-nucleon distortions a newly
developed realistic potential within the Bonn family of
potentials, valid well above the pion production threshold, is
considered.
In the $\pi^+$ production case, the $NN$ potential, the $\pi N$
relative $p$ waves and the treatment of the exchanged pion energy
build up the observed cross-section strength.
\end{abstract}
\maketitle
%
%
%
A central problem underlying the non-relativistic
quantum-mechanical operator for any mechanism of pion production
in the framework of time-ordered perturbation theory (TOPT) is
that the final- and initial-state interaction diagrams correspond
to different off-energy shell extensions of the amplitudes, since
energy is not conserved at individual vertices.
This problem does not occur in the S-matrix
approach\cite{Chemtob:1971pu} which was applied to derive the
important Z-diagram operators\cite{Lee:1993xh}. Moreover, it was
recently shown that this approach with a particular prescription
for the energy of the exchanged pion reproduces well the energy
dependent re-scattering operator for neutral pion production
resulting from TOPT\cite{Malafaia:2004cu}.

We report here that a calculation using the S-matrix approach with
a $NN$ potential tuned above pion production and a sufficient
number of $3$-body channels with $\pi N$ relative $p$-waves,
overcomes the problem of insufficient production strength in the
$\pi^+$ charge channel found in \rf{daRocha:1999dm}.
Therefore, this work shows that it is possible to have an overall
description of all the charge channels from the same underlying
physical processes and realistic sub-cluster interactions. As in
\rf{daRocha:1999dm}, and alternatively to the J\"{u}lich
group\cite{Hanhart:1995ut}, we take the $\chi$PT $\pi$N
amplitude\cite{Hanhart:2002bu} to construct the irreducible pion
re-scattering term. Together with this contribution, we consider
the direct-production term, $Z$-diagrams mediated by scalar and
vector exchanges, and the explicit $\Delta$-isobar excitation.

For the initial state, the nucleon-nucleon interaction is
necessarily needed above the pion production threshold. We apply
here for the first time an $NN$ interaction recently developed by
the Ohio group, including two boson exchange box contributions,
with Delta$(1232)$ and N*$(1440)$ intermediate
states\cite{Elster:1988zu}.
As an extension of the family of the Bonn potentials, it describes
well the $NN$ phase-shifts and inelasticities up to $1$~GeV.
%
%
%
%

\begin{figure}
\includegraphics[width=.146\textwidth,keepaspectratio]{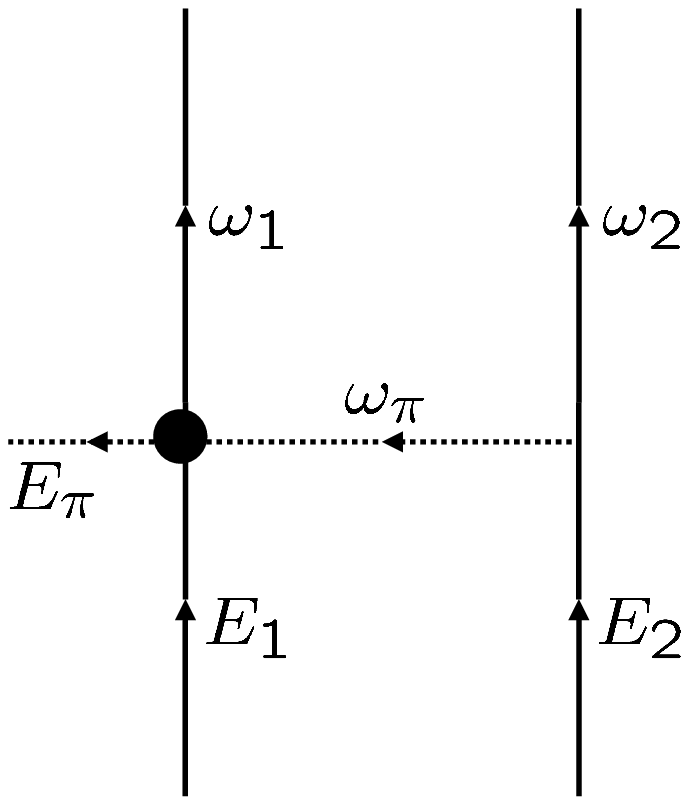}
\caption{The re-scattering diagram. \label{diagvresc}}
\end{figure}

In the S-matrix prescription, the effective operators are defined
only on-energy-shell through the non-relativistic reduction of the
corresponding Feynman diagrams where the energy is conserved at
each vertex\cite{Chemtob:1971pu}. All nuclear currents and other
transition operators are defined to be consistent with an
hermitian energy independent $NN$ potential.
For the pion re-scattering diagram (\fig{diagvresc}), the S-matrix
prescription leads to a single effective operator (both for FSI
and ISI diagrams) of the form\cite{Malafaia:2004cu}:
\begin{equation}
 \hat{O}_{rs}^{S}=
 \frac{f(\Omega)}{(\Omega)^2-\left(m_\pi^2+ \vec{q}^{\, \prime 2}\right)} \, ,
\label{msmat}
\end{equation}
where, adopting the notation of \rf{Malafaia:2003wx}, $\vec{q}^{\,
\prime}$ is the momentum of the exchanged pion and $f(\Omega)$ is
the product of the $\pi N$ amplitude with the $\pi NN$ vertex.
Whenever the operators are used in convolution integrals for the
distortion of the nucleonic states, the momentum variable is free.
The symmetric implementation of the energy conservation condition
$\omega_\pi=E_2-\omega_2=-\left(E_1-\omega_1-E_\pi \right)$ at the
vertices of \fig{diagvresc} is then used
\begin{equation}
\Omega=\frac{\left(E_2-\omega_2\right)}{2}+\frac{\left(\omega_1+E_\pi-E_{1}
\right)}{2}.
\end{equation}
This prescription minimizes the deviation between the S-matrix and
the TOPT results for the re-scattering
diagram\cite{Malafaia:2004cu}.
%
%
%

The explicit $\Delta$-isobar propagation contribution is
\begin{eqnarray}
V_{\pi \pi N \Delta} &=& i \left(\frac{h_{A}}{2 f_{\pi}}
\right)^{2} \frac{4g_{A}}{18f_{\pi }} \frac{1}{\Delta-E_{\pi}}\frac{1}{\Omega^{2}-\omega_{\pi}^{2}}\label{opdelta} \\
&&\left[ 2 \tau_{a}^{\left( 2 \right)}+\tau_{b}^{\left( 1 \right)}
\tau_{c}^{\left(2 \right)}\varepsilon^{abc} \right]\left(
\vec{q}_{\pi} \cdot \vec{q}^{\prime} \right) \vec{\sigma}^{\left(
2 \right)} \cdot \vec{q}^{\prime} \nonumber
\end{eqnarray}
%
Since the $\Delta$ is added explicitly to the $\pi N$ amplitude,
the static limit of \eq{opdelta} is subtracted from the $\chi$PT
amplitude\cite{daRocha:1999dm}. This redefines the $c_{3}=-5.29
\ugev^{-1}$ parameter to $c_{3}^{\prime}=c_{3}-c_{3}^{\Delta}$,
with $c_{3}^{\Delta}\approx-h_{A}^{2}/\left(18
m_{\pi}\right)=-2.78\ugev^{-1}$.
%
%
%
%
To go beyond threshold, we include $p$-wave pion production and
$NN$ partial waves with high relative angular momentum.
Details will be given in \rf{MPEA}.
%
%
%
%

\begin{figure*}
\begin{center}
\includegraphics[width=.99\textwidth,keepaspectratio]{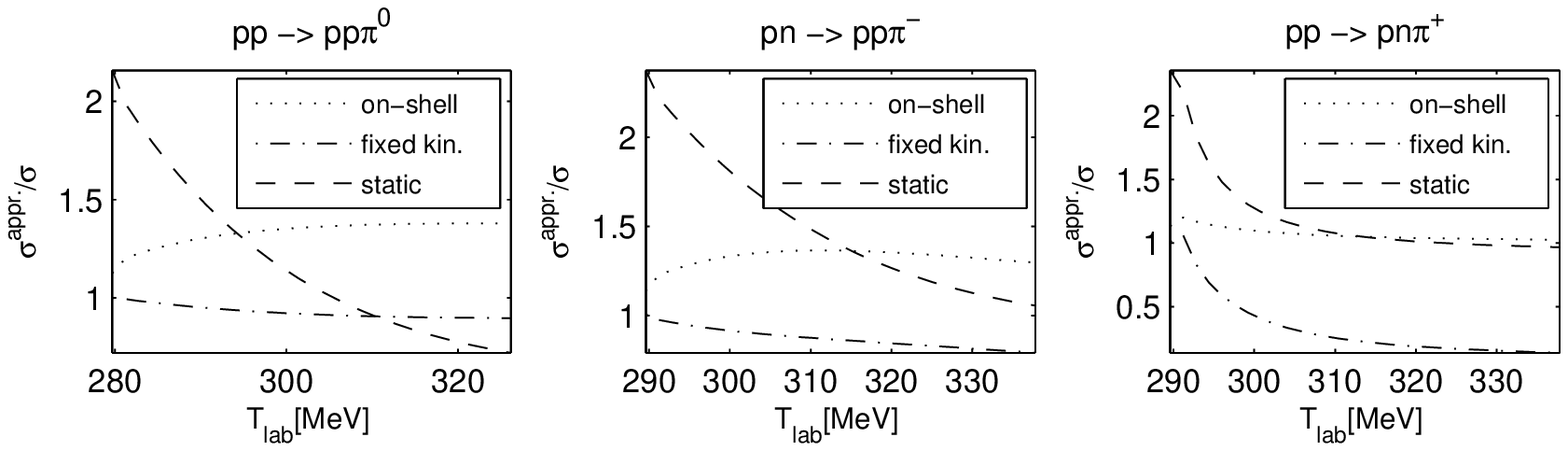}
  \caption{Ratio between the cross section with different approximations and the S-matrix cross section.
   The dotted, dashed-dotted and dashed line are, respectively, the on-shell approximation,
   fixed-kinematics and static approximation.} \label{pifcptbepr}
\end{center}
\end{figure*}

We compare in \fig{pifcptbepr} the cross sections calculated
within the S-matrix approach to those obtained with frequently
used approximations. For all the charge channels, the deviation is
largest for the static approximation ($\Omega=0$), which
overestimates the cross section by a factor of $2$ (close to
threshold).
The on-shell approximation ($\Omega=E_{2}-\omega_{2}$, difference
between the nucleon on-shell energies before and after pion
emission) also deviates from the reference result. It may be off
$20 \%$ for all the charge channels, even close to threshold. For
the $\pi^{0}$ and $\pi^-$ production cases, the same happens for
the fixed kinematics approximation ($\Omega=m_{\pi}/2$).
For $pp \rightarrow pn\pi^{+}$, this approximation underestimates
the cross section by a factor of $1.5-2.5$ near threshold and $5$
at higher energies. This significant difference arises from the
$\pi N$ relative $p$-waves being very sensitive to the energy
prescription. They are particularly important for the $\pi^{+}$
channel and magnify at higher energies the difference between the
S-matrix approach and the fixed-kinematics approximation.
The results for $\pi^{+}$ give insight into the discrepancy with
the experimental data by factors $2$-$5$ reported in
\rf{daRocha:1999dm}, where the calculation also used the $\chi$PT
$\pi$N amplitude as here, but included only relative $\pi N$
$s$-wave states and assumed fixed kinematics.
%
%
%
%

\begin{figure*}
\begin{center}
\includegraphics[width=.77\textwidth,keepaspectratio]{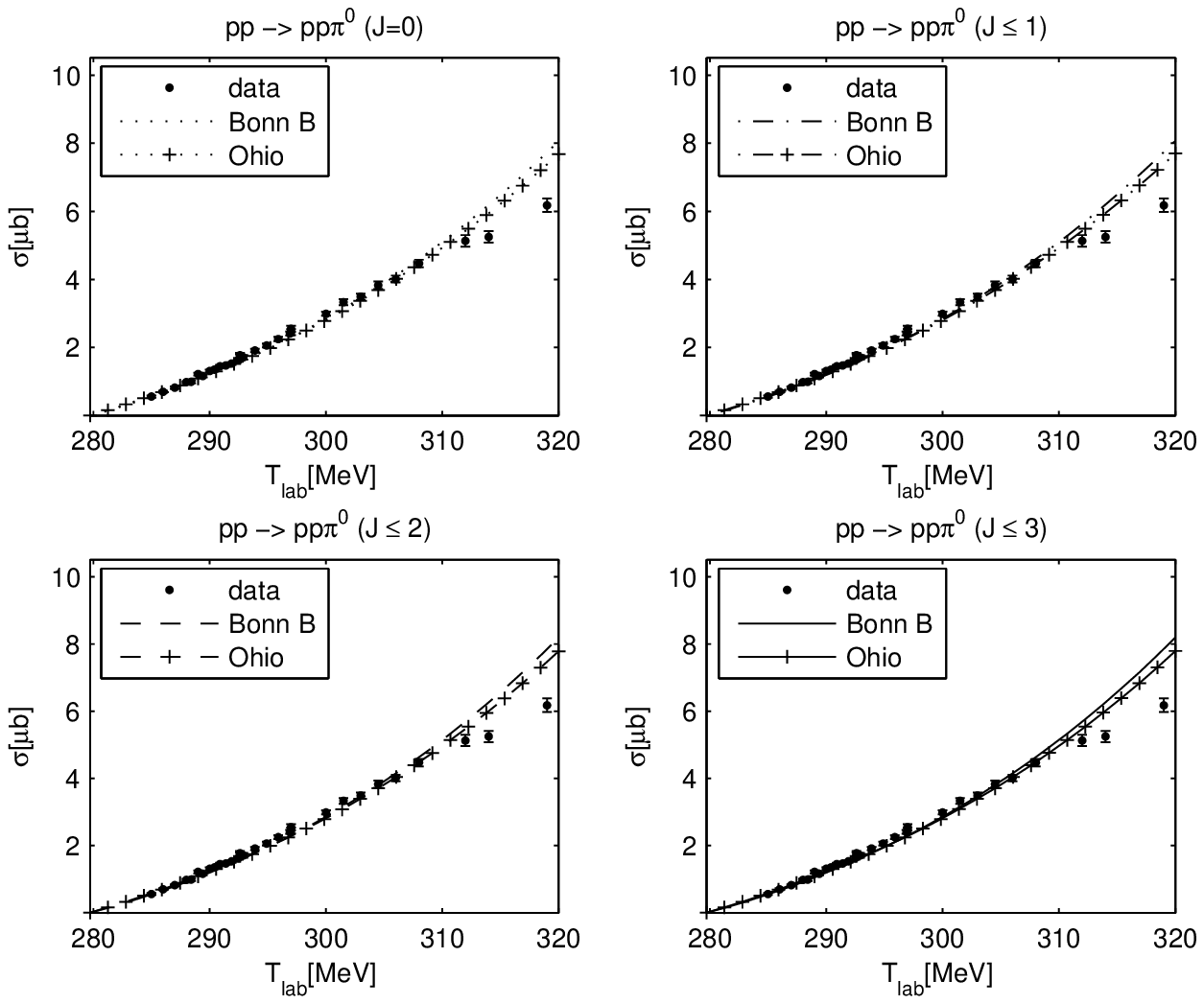}
 \caption{Effect of the
   $NN$ interaction on the $pp
   \rightarrow pp \pi^{0}$ cross section. Dotted ($J=0$), dashed-dotted ($J \leq 1$), dashed
   ($J \leq 2$) and solid ($J \leq 3$) lines with (without) +'s
   correspond to the Ohio (Bonn B) potential.
   The data points are from \rf{Meyer:1992jt}.\label{pi0cmcp}}
\end{center}
\end{figure*}

\begin{figure*}
\begin{center}
\includegraphics[width=.77\textwidth,keepaspectratio]{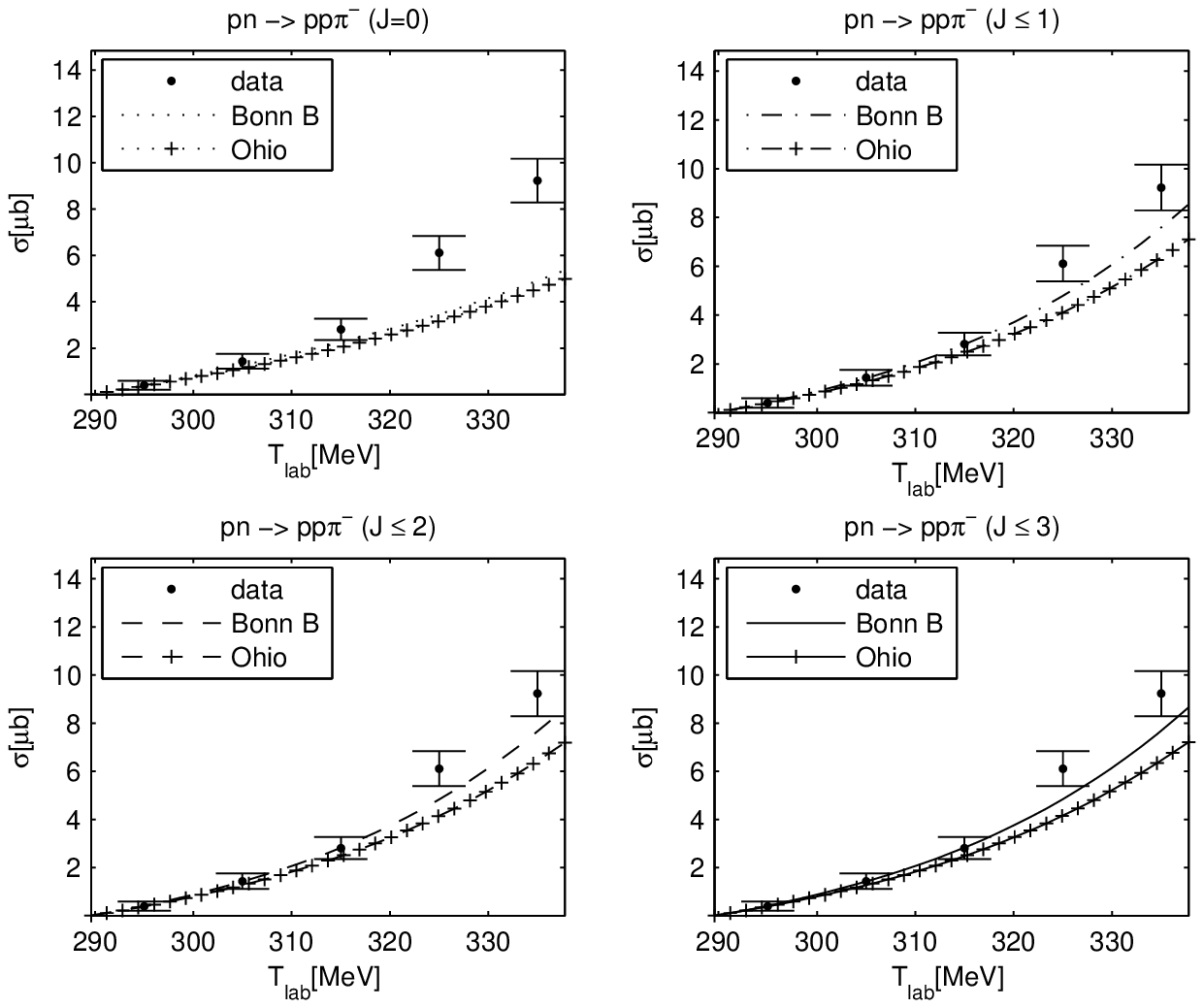}\\
  \caption{The same of \fig{pi0cmcp} but for the $pn
   \rightarrow pp \pi^{-}$ cross section. The data points are from
   \rf{Bachman:1995gn}. \label{pimcmcp}}
\end{center}
\end{figure*}

\begin{figure*}
\begin{center}
\includegraphics[width=.77\textwidth,keepaspectratio]{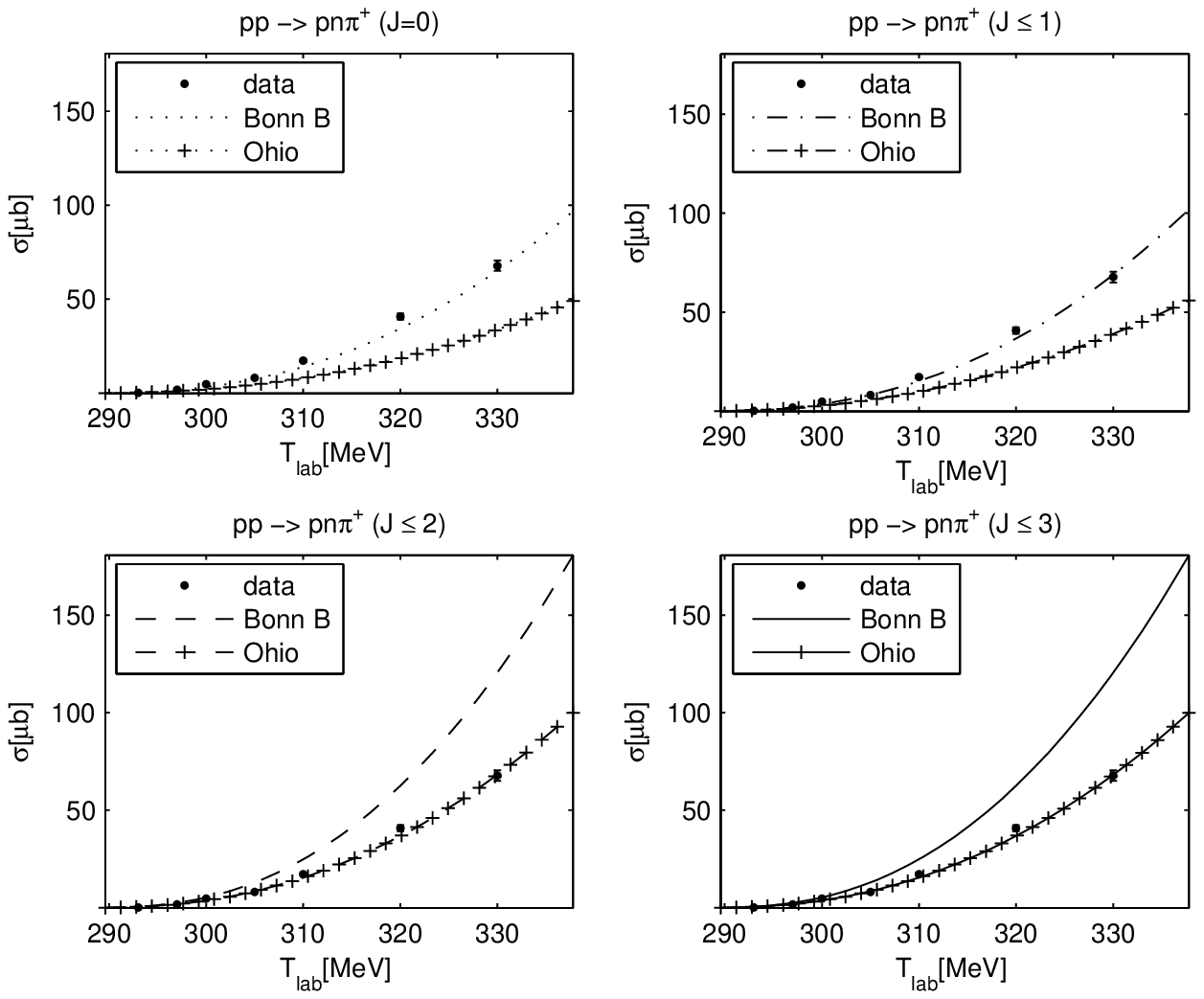}
  \caption{The same of \fig{pi0cmcp} but for the $pp \rightarrow pn \pi^{+}$
    cross section. The data points are from \rf{Hardie:1997mg}. \label{pipcmcp}}
\end{center}
\end{figure*}

We compare the cross sections with the Ohio and the Bonn B
potentials, and simultaneously illustrate the convergence of the
amplitude partial waves with increasing angular momentum $J$, in
\figs{pi0cmcp}{pipcmcp}. As expected, the importance of channels
with high $J$ increases with increasing laboratory energy
$T_{lab}$.
For $\pi^0$ production, the $J=0$ channel alone describes the data
and suffices for convergence (\fig{pi0cmcp}).
For $\pi^{-}$ production (\fig{pimcmcp}) however the $J=1$
channels are needed.
For $\pi^{+}$ production (\fig{pipcmcp}), the convergence is
slower than in the other two reactions, as the $J=2$ channels are
needed for convergence.
%


We further verified that the direct-production and re-scattering
mechanisms alone are not sufficient to describe the $\pi^{0}$
production data, and Z-diagrams are decisive, as known.
For $\pi^{-}$ production, the Weinberg-Tomozawa term, which does
not contribute to $\pi^{0}$ production, is important relative to
the direct production term. The Z-diagrams have also an important
role.
For $\pi^{+}$ production, the Ohio $NN$ model gives a
significantly better description of the data. It should be noticed
that in contrast to the Bonn B potential, it includes explicit
$\Delta$ contributions. For $\pi^{+}$ production, the  $\Delta$
contributions dominate and increase with energy.
The general trends obtained here were also found for the
J\"{u}lich phenomenological model\cite{{Hanhart:1995ut}}. There
all short range mechanisms are included through $\omega$-exchange
and adjusted to reproduce the total $pp \rightarrow pp \pi^{0}$
cross section close to threshold. In our calculation no adjustment
is made. The parameters for the $Z$-diagrams and for the $\Delta$
contribution are taken from the $NN$ interaction employed.
%
%
%

We achieved an overall S-matrix description of the cross section
both for charged and neutral pion production, from the same
underlying processes and interactions.
The crucial role of the $\Delta$ in $\pi^{+}$ production enhances
the importance of the $\pi N$ $p$-wave states resulting in a
slower convergence and in a higher sensitivity to the fixed
kinematics approximation.
By taking these channels and simultaneously going beyond the fixed
kinematics approach, we obtained a good description of the $pp
\rightarrow pn \pi^{+}$ reaction, problematic until
now\cite{daRocha:1999dm} when the $\chi$PT $\pi N$ amplitude is
used.
The newly developed Ohio interaction tuned above the pion
production threshold furthermore improves the description of the
$pp \rightarrow pn \pi^{+}$ reaction.
Our results should be relevant for high precision calculations, as
in charge symmetry breaking studies.
\begin{acknowledgments}
V.M. was supported by FCT under the grant SFRD/BD/4876/2001,
M.T.P. was supported by the grant POCTI/FNU/50358/2002 and J.A.
was supported by the grant  GA CR 202/03/0210. Ch.~E~.
acknowledges the support of the U.S. Department of Energy under
contract No. DE-FG02-93ER40756 with Ohio University.
\end{acknowledgments}



\end{document}